\documentclass[aps, onecolumn, amsmath, amssymb]{revtex4-1}
\usepackage{dcolumn}% Align table columns on decimal point
\usepackage{bm}% bold math
\usepackage{amsmath}
\usepackage{amsfonts}
\usepackage{ulem}
\usepackage[dvips]{graphicx}
\usepackage{multirow}
\usepackage{subfigure}
\usepackage{times}
\usepackage{setspace}
\usepackage{color}
\usepackage[raggedright]{titlesec}

\begin{document}

\title{Cascading Failures in Bi-partite Graphs: \\Model for Systemic
  Risk Propagation }

\author{Xuqing Huang$^1$}
\author{Irena Vodenska$^{1,2}$}
\email{ Correspondence and requests for materials should be 
addressed to I. V. (vodenska@bu.edu).}
\author{Shlomo Havlin$^{1,3}$} 
\author{H. Eugene Stanley$^1$}

\today

\affiliation{$^1$Center for Polymer Studies and Department of Physics,\\
Boston University, Boston, Massachusetts 02215 USA\\
$^2$Administrative Sciences Department, 
Metropolitan College, Boston University, Boston, Massachusetts 02215 USA\\
$^3$Department of Physics, Bar-Ilan University, Ramat-Gan 52900, Israel
}

\begin{abstract} 
\noindent
 
As economic entities become increasingly interconnected, a shock in a 
financial network can provoke significant cascading failures throughout the system. 
To study the  systemic risk of financial systems, we create a bi-partite banking 
network model composed of banks and bank assets and propose a cascading failure model to describe the risk propagation process during crises.
We empirically test the model with 2007 US commercial banks balance sheet data 
and compare the model prediction of the failed banks with the real failed
banks after 2007. We find that our model efficiently identifies a significant portion of the actual failed banks reported by Federal Deposit Insurance Corporation. The results suggest that this model could be useful for systemic risk stress testing for financial systems.
The model also identifies that commercial rather than residential real estate assets are major culprits for the failure of over 350 US commercial banks during  2008-2011.

\end{abstract}

\maketitle

There have been dramatic advances in the field of complex networks in
recent years \cite{Watts1998, barabasi1999, Alber2002, Newman2006,
  Cohen2010, Buldyrev2010}. The Internet, airline routes and electric
power grids are all examples of networks in which connectivity between
network components is essential. 

Because of the strong connectivity, 
catastrophic cascading failure of nodes in networks can happen when the system 
is under a shock, especially if the shocked nodes represent hubs, or have 
high centrality measures in the network~\cite{Cohen2001, Motter2002, 
Smart2008, Vespignani2010, Huang2011}. 
So, in order to minimize the systemic risk, these networks should be 
designed to be robust to external shocks. 
In the wake of the recent global financial crisis, increased attention 
has been given to the study of the dynamics of economic systems and to systemic 
risk in particular.
The widespread impact of the current EU sovereign debt crisis and the 2008
world financial crisis show that as economic systems become increasingly interconnected, 
local exogenous or endogenous shocks can provoke global cascading system failure that is
difficult to reverse and that cripples the system for a prolonged period
of time. Thus policy makers are compelled to create and implement safety measures
that can prevent cascading system failures or soften their systemic impact. 
Based on the success of complex networks in modeling interconnected systems, 
applying complex network theory to study economical systems has been under the spot light
\cite{May2008, Garas2010, Johnson2011, Haldane2011, Schweitzer2012, Battiston2012}.

There are two channels of risk contagion in the banking system, (i)
direct interbank liability linkages between financial institutions and
(ii) contagion via changes in bank asset values.  The former, which has
been given extensive empirical and theoretical study \cite{Wells2002,
Furfine2003, Upper2004, Elsinger2006, Nier2007}, focuses on
the dynamics of loss propagation via the complex network of
direct counterpart exposures following an initial default. The latter,
based on bank financial statements and financial ratio analysis, has 
received scant attention. A financial shock that contributes to the bankruptcy 
of a bank in a complex network will cause the bank to sell its assets. 
If the market's ability to absorb these sales is less than perfect, 
the market prices of the assets that the bankrupted bank sells will decrease.
Other banks that own similar assets could also fail because of loss in 
asset value and increased inability to meet liability obligations.
This imposes further downward pressure on asset values and contributes 
to further asset devaluation in the market. Damage in the banking network thus 
continues to spread, and the result is a cascading of risk propagation 
throughout the system \cite{Cifuentes2005, Tsatskis2012}.
In this paper we model the risk contagion via changes in asset values in 
the banking system.

In the past 2008 financial crisis, 371 commercial banks failed 
between 1/1/2008 and 7/1/2011. The Failed Bank Lists from the Federal Deposit 
Insurance Corporation (FBL-FDIC) records the names of failed banks and 
the time when the banks failed. We use this list as an experimental
benchmark to our model. The other dataset that we use is the US Commercial 
Banks Balance Sheet Data (CBBSD) from Wharton Research Data Services, 
which contains the amounts of assets in each category that the US 
commercial banks had on their balance sheets (see Method section for more detail). 
We use this dataset as an input to our model. 

The contributions of this paper are as follows. We first 
analyse the properties of the failed banks from FBL-FDIC, examining the 
weights in specific assets as well as equity to asset ratios.
We then construct a bipartite banking network that is composed of two types of nodes, 
(i) banks and (ii) bank assets. Link between a bank and a bank asset exists when 
the bank has the asset on its balance sheet. We also develop a cascading failure 
model to simulate the crisis spreading process in the bipartite network. 
We then populate the model by the banks' balance sheet data (CBBSD) for 2007, 
and run the cascading failure model by initially introducing a shock to the banking system. 
We compared the failed banks identified by model with the actual failed banks from the
FBL-FDIC from 2008 to 2011, and find that our model simulates well the crisis spreading process
and identifies a significant portion of the actual failed banks. Thus, we suggest 
that our model could be useful to stress test systemic risk of the banking system.
For example, we can test each particular asset or groups of assets influence on the 
overall financial system i.e. if the agricultural assets drop by $20\%$ in value, we can study which banks could 
be vulnerable to failure, and offer policy suggestions to prevent such failure, 
such as requirement to reduce exposure to agricultural loans or closely monitor 
the exposed banks. 
Finally, we show that sharp transition can occur in the model as 
parameters change. 
The bank network can switch between two distinct regions, stable
and unstable, which means that the banking system can either survive and
be healthy or completely collapse. Because it is important that policy
makers keep the world economic system in the stable region, we suggest
that our model for systemic risk propagation might also be applicable to
other complex financial systems, e.g., to model how sovereign debt value
deterioration affects the global banking system or how the depreciation
or appreciation of certain currencies impact the world economy.

\section*{Results}

\begin{figure}[h!]
  \centering 	
  \includegraphics[width=0.8\textwidth ]{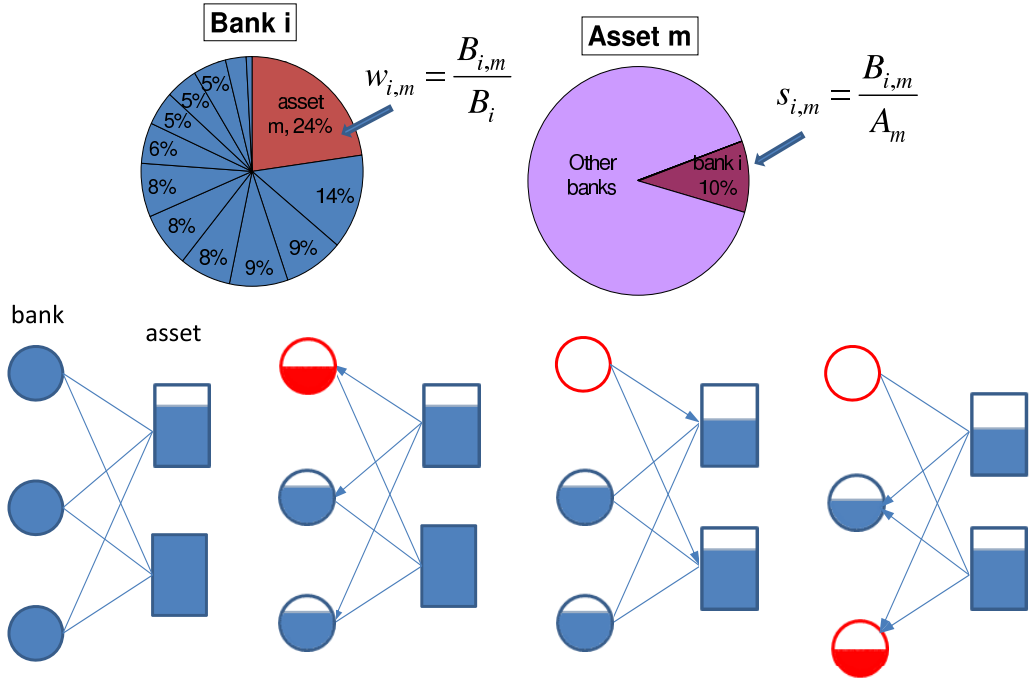}
  \caption{
  Bank-asset bipartite network model with banks as one node type and assets as the other node type. Link between a bank and an asset exists if the bank has the asset on its balance sheet. Upper panel: illustration of bank-node and asset-node. $B_{i,m}$ is the amount of asset $m$ that bank $i$ owns. Thus, a bank $i$ with total asset value $B_i$ has $w_{i,m}$ fraction of its total asset value in asset $m$. $s_{i,m}$ is the fraction of asset $m$ that the bank holds out. Lower panel: illustration of the cascading failure process. The rectangles represent the assets and the circles represent the banks. From left to right, initially, an asset suffers loss in value which causes all the related banks' total assets to shrink. When a bank's remaining asset value is below certain threshold (e.g. the bank's total liability), the bank fails. Failure of the bank elicits disposal of bank assets which further affects the market value of the assets. This adversely affects other banks that hold this asset and the total value of their assets may drop below the threshold which may result in further bank failures. This cascading failure process propagates back and forth between banks and assets until no more banks fail.
}
\label{interbank_model}
\end{figure}

\noindent{\bf Properties of failed banks.}  
To build a sound banking system network and systemic risk 
cascading failure model, we need to study the properties of the failed banks.
The asset portfolios of commercial banks contain such asset categories as commercial loans,
residential mortgages, and short and long-term investments.
We model banks according to how they construct their asset portfolios
(upper panel of fig.~\ref{interbank_model}). For each bank, the CBBSD
contains 13 different non-overlapping asset categories, e.g., bank $i$
owns amounts $B_{i,0}, B_{i,1}, ... , B_{i,12}$ of each asset,
respectively.  The total asset value $B_i$ and total liability value $L_i$ 
of a bank $i$ are obtained from CBBSD dataset.
The weight of each asset $m$ in the overall asset portfolio of a bank $i$
is then defined as $w_{i,m}\equiv B_{i,m}/B_i$.  From the perspective of
the asset categories, we define the {\it total market value\/} of an
asset $m$ as $A_m\equiv \sum_iB_{i,m}$. Thus the market share of bank $i$ in
asset $m$ is $s_{i,m}\equiv B_{i,m}/A_m$.

\begin{figure}[h!]
  \centering 	
  \includegraphics[width=\textwidth ]{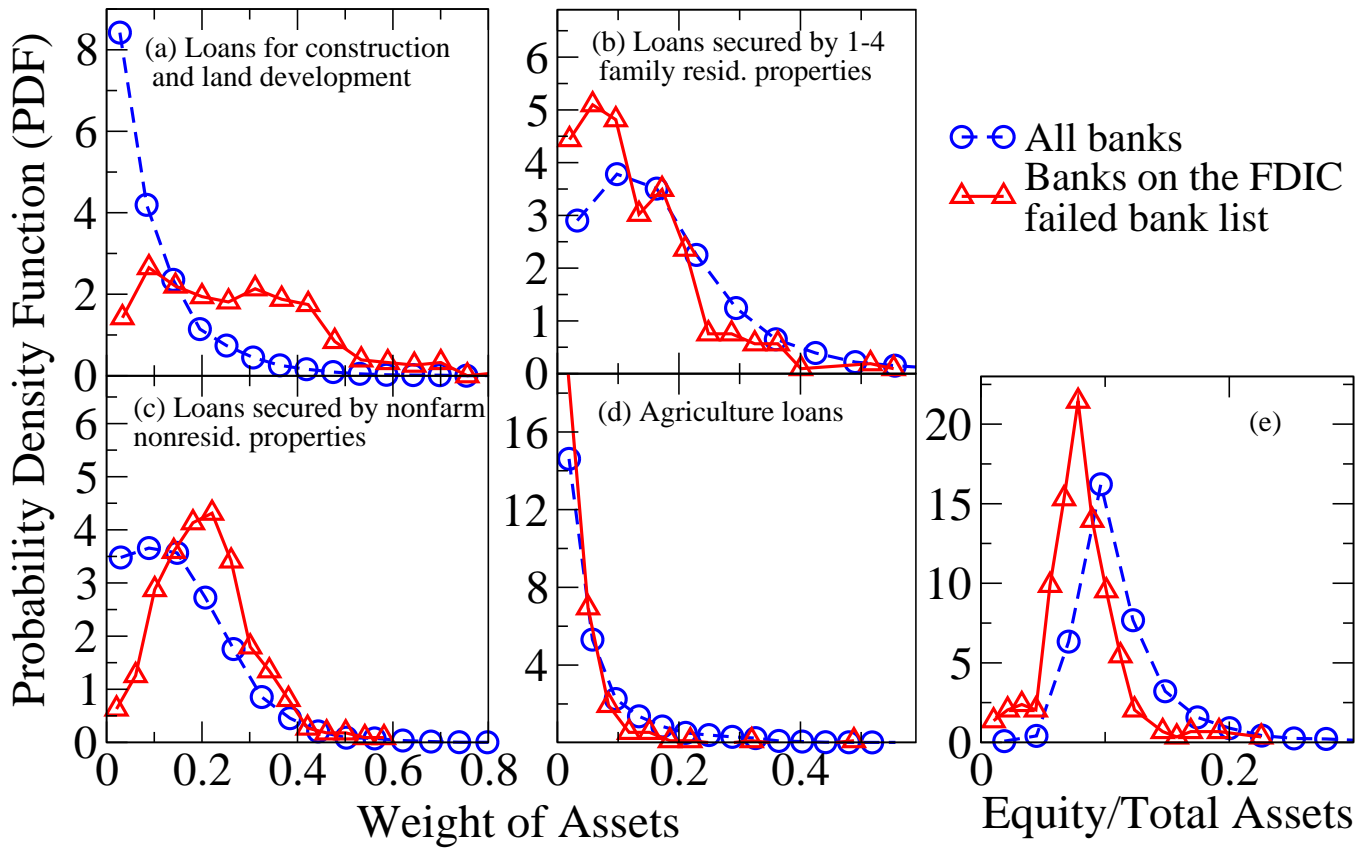}
  \caption{
  Comparison of probability density functions (PDF) of weight of typical assets and equity ratios between all banks and FDIC listed failed banks for 2007. (a) PDF of the weight of loans for construction and land development in banks' total asset. (b) PDF of the weight of loans secured by 1-4 family residential properties in banks' total assets. (c) PDF of the weight of loans secured by nonfarm nonresidential properties in banks' total assets. (d) PDF of the weight of agriculture loans in banks' total assets. (e) PDF of banks' equity to asset ratios. Blue circles curves represents PDFs of all banks, red triangles represents PDFs of those banks that are on the FDIC failed bank list.  
  }
  \label{pdf}
\end{figure}

To study the properties of failed banks between 2008 and 2011, we focus
on the weight of each bank's assets. For certain assets, we find that
the asset weight distributions for all banks differ from 
the asset weight distributions for failed banks. 
Figures~\ref{pdf}(a) and \ref{pdf}(c) show that, unlike 
survived banks, failed banks cluster in a region heavily weighted 
with construction and development loans and loans secured by 
nonfarm nonresidential properties. Failed
banks have less agricultural loans in their asset portfolios compared to
survived banks (fig.~\ref{pdf}(d)). These results confirm the
nature of the most recent financial crisis of 2007--2011 in which bank
failures were largely caused by real estate-based loans, including loans
for construction and land development and loans secured by nonfarm
nonresidential properties~\cite{Cole2011}.  In this kind of financial
crisis, banks with greater agricultural loan assets are more financially
robust \cite{Corner2011}. Figure~\ref{pdf}(e) shows that failed banks 
tend to have lower equity to asset ratios,
i.e. failed banks generally had higher leverage ratios than survived 
banks during the financial crisis of 2008-2011 ~\cite{Gopalan2010}.

\noindent{\bf Cascading failure propagation model.} To study the
systemic risk of the banking system as complex networks, we construct 
a cascading failure model based on the facts presented in the previous section.

We first build a bipartite network which contains two 
types of nodes, banks on one hand and bank assets on the other. 
Link exists between a bank and an asset when the bank 
has the asset on its balance sheet. No links between banks or between 
assets exist. 
To simulate the cascading failure process, we develop and apply the 
following model with three parameters $p$, $\eta$ and $\alpha$
(illustrated in fig.~\ref{interbank_model}):
\begin{enumerate}
\item We initially shock certain asset $m$, reducing the 
{\it Total Market Value } of asset $m$ to $p$ fraction of its
original value, $p \in [0,1]$. The smaller the $p$ is, the larger the shock.
When $p$ is 0, the total market value of asset $m$ is wiped out.
When $p$ is 1, no shock is imposed. 

\item When the market deteriorates, each bank $i$ that owns the shocked
asset $m$ will experience $B_{i,m}(1-p)$ reduction in 
value, where $B_{i,m}$ is the amount of asset $m$ that is on bank $i$'s
balance sheet.

\item When the total asset value of a bank declines to a level below
the level of promised payments on the debt, it causes distress or default.
The total asset value that triggers an incidence of distress lies
somewhere between the book value of total liabilities and short-term liabilities.
In the corporate sector default analysis, Moody's Analytics 
used the sum of short-term debt, interest payments and half of long-term 
debt~\cite{Merton1974, KMV2003, Gray2007} as the distress barrier.
However, in the past financial crisis, external aid from other financial 
institutes or from the government played a significant role in distorting this distress barrier,
thus even when a bank's total value of assets was below its liabilities, 
the bank could still survive. We describe these combined
effects using random number $r$ that is uniformly distributed in range $[0, \eta]$,
where $\eta \in [0,0.5]$ is a parameter controlling tolerance of a bank's assets
being below its liabilities.
We define the distress barrier to be $(1-r)\cdot L_i$, such that a bank fails when 
$B_i < (1-r)\cdot L_i$.
For such distress barrier with evolving randomness, the probability $P_(B_i, L_i)$ 
for a bank $i$ to fail can be written as
\begin{equation}
P(B_i, L_i) = \left\{ \begin{array}{ll}
 0 &\mbox{ if $B_i \geq L_i$} \\
 (L_i-B_i)/(\eta L_i) &\mbox{ if $\eta \neq 0$, $L_i > B_i > (1-\eta)L_i$ } \\
 1 &\mbox{ if $ (1- \eta)L_i > B_i $ }
       \end{array} \right.
\end{equation}

\item We assume that when a bank $i$ fails, the overall market value of each 
asset $m$ that the failed bank owns suffers $\alpha B_{i,m}$ value deduction, 
where $\alpha \in [0, 1]$ is a third parameter in the model that describes 
the market's reaction to a bank failure. The unit price
of asset $m$ becomes $\frac{A_m - \alpha B_{i,m}}{A_m}$ of its original price.
That is because the failed banks need to sell assets to meet their 
liabilities and the market's ability to absorb this sale is not
perfect, which leads to price decrease of the affected assets.
The loss of the market value of each asset $m$ is proportional
to $B_{i,m}$, the amount of asset $m$ that the failed bank $i$ owns.
Depending on the liquidity of an asset, $\alpha$ can be between 0 and 1. When
an asset is extremely liquid, the market value of the asset will not be adversely 
affected by asset sales, $\alpha = 0$. When the market is extremely illiquid,
then the value of asset could potentially have zero value. 
Thus the aggregated total market value of asset $A_m$ will be reduced to 
$A_{m}-B_{i,m}$, which corresponds to $\alpha=1$.

\item Further deterioration of asset values can then contribute to
failure of more banks.
Thus the damage in the bipartite network spreads between banks and assets 
bidirectionally until the cascading failure stops.
\end{enumerate}

Usually financial crises start with a burst of economic bubbles.
The correspondence of the model's initial shock parameter $p$ in reality 
can be described as the drop of certain asset value at the beginning of a crisis.
For example, when the dot-com bubble burst, the technology 
heavy NASDAQ Composite index lost $66\%$ percents of its value, plunging 
from the peak of 5048 in March 10, 2000 to the 1720 in April 2, 2001.

\begin{figure}[h!]
  \centering 	
  \includegraphics[width=\textwidth ]{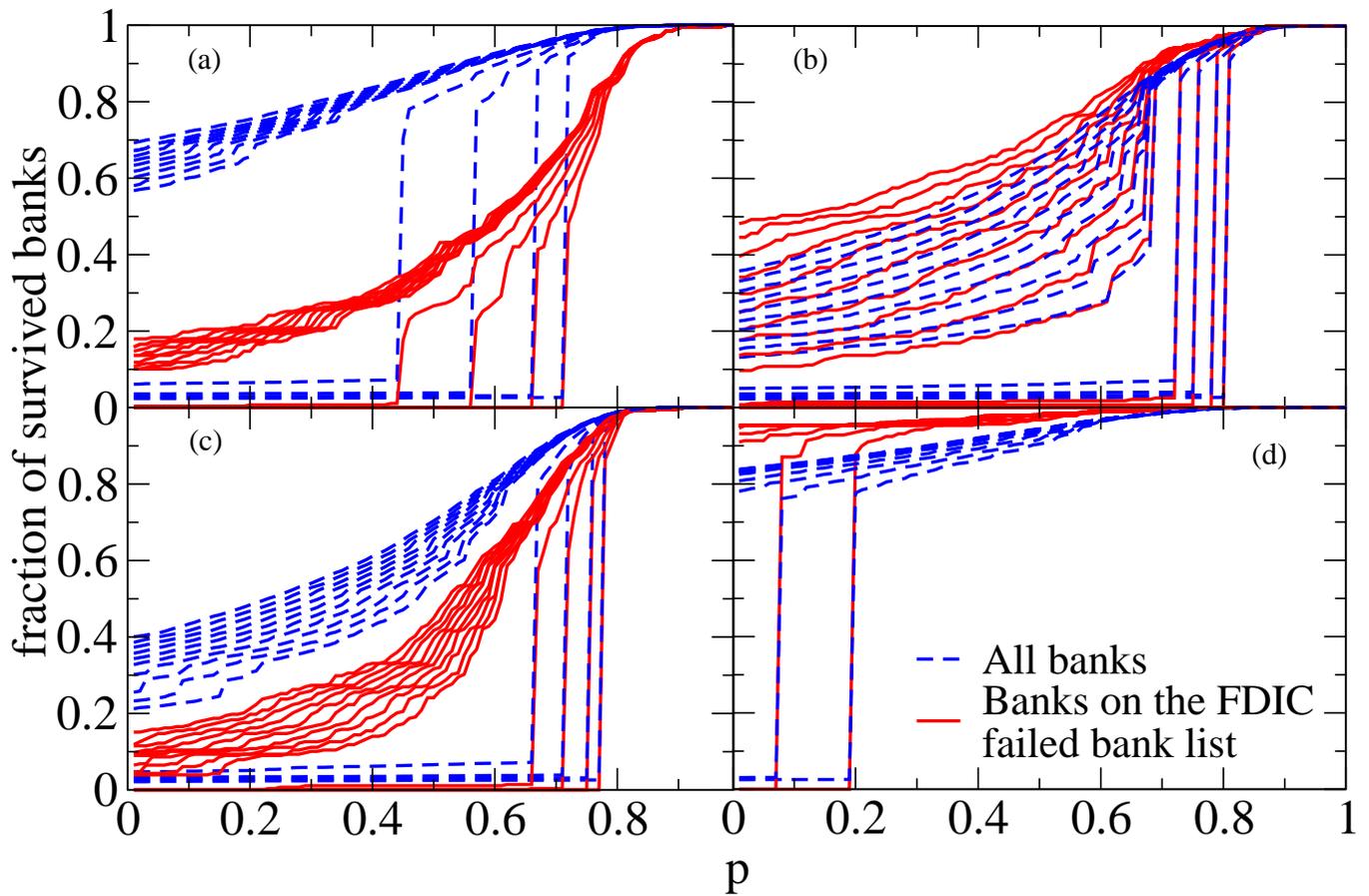}
  \caption{
  Fraction of survived banks after cascading failures as function of the initial loss of value of certain asset, with $\eta = 0$. Blue dashed lines represent the fraction of survived banks out of all banks, and the red solid lines represent the fraction of survived banks out of the 278 failed banks from FDIC failed bank list. The parameter $\alpha$ is changed from $0$ to $0.1$ by $0.01$ to produce 10 lines for each case. (a) Initial shock is imposed to loans for construction and land development. The red solid lines are significantly lower than the blue dashed lines separating clearly the failed banks from the set of all banks. (b) Initial shock is imposed to loans secured by 1-4 family resid. properties. The red solid lines and blue dashed lines are entangled. (c) Initial shock on loans secured by nonfarm nonresid. properties. The red solid lines are distinguishably lower than the blue dashed lines, similarly as in the case under (a). (d) Initial shock on agricultural loans. The red solid lines are slightly higher than the blue dashed line, not showing clear distinction between failed and non failed banks.
  }
  \label{percolation}
\end{figure}

\noindent{\bf Empirical test and analysis.} 
To empirically test our model, we introduce a shock into the banking system by
reducing $(1-p)$ percentage of the value of a single asset $m$. 
We then monitor the progression of bank failure until the cascading process stops. 
We examine two distinct groups of banks 1) all the analyzed banks from CBBSD dataset, 
and 2) the banks from the FDL-FDIC failed bank list. We then study the fraction of banks 
that were identified as survived by our model in both groups.
We plot both of these fractions versus the sizes
of initial shocks in fig.~\ref{percolation}, for parameter $\eta=0$. 
The four plots correspond to four typical assets being initially 
shocked respectively.
Figure~\ref{percolation}(a) and figure~\ref{percolation}(c) show that when
the commercial real estate loans, i.e. loans for construction and land 
development and loans secured by nonfarm nonresidential properties, 
suffer initial shock respectively, the survival rate of the banks
from the first group (all banks), according to our model, is distinctly 
above the survival rate of the second group of banks (FBL-FDIC failed banks list).
This illustrates that when the commercial real estate loans are 
initially shocked, the model can identify the actual failed banks efficiently.
Figures~\ref{percolation}(b) and \ref{percolation}(d) show that
when we impose initial shock on loans secured by 1-4 family residual properties 
or agricultural loans, the model does not clearly separate the two groups of banks.
This result indicates that the commercial bank failures during the 2008 financial 
crisis stems from value deterioration of commercial real estate loans.

To quantitatively test the efficiency of the model in identifying failed banks,
we use the receiver-operating-characteristic (ROC) curve analysis, which plots 
the fraction of true positives out of the positives and the fraction 
of false positive out of the negatives for a binary classifier system.
ROC curve analysis is a standard method in signal detection theory as
well as in psychology, medicine and biometrics~\cite{ROC_curve}. 
We choose a parameter combination of $p$, $\eta$ and $\alpha$ to
run the model to determine which banks fail, and compare this prediction 
with the FDIC list of failed banks.
The true positive rate is defined as the fraction of the actual failed 
banks that are also identified as failed in our model.
The false positive rate is the fraction of banks that are not on the FDIC
list of failed banks but are identified as failing by our model. Each point
in the ROC curve corresponds to one parameter combination. 
A complete random guess would give points along the diagonal line from the 
left bottom to the top right corner. The more a point is above the diagonal
line, the stronger predictive power the model has.

\begin{figure}[h!]
  \centering 
  \includegraphics[width=\textwidth ]{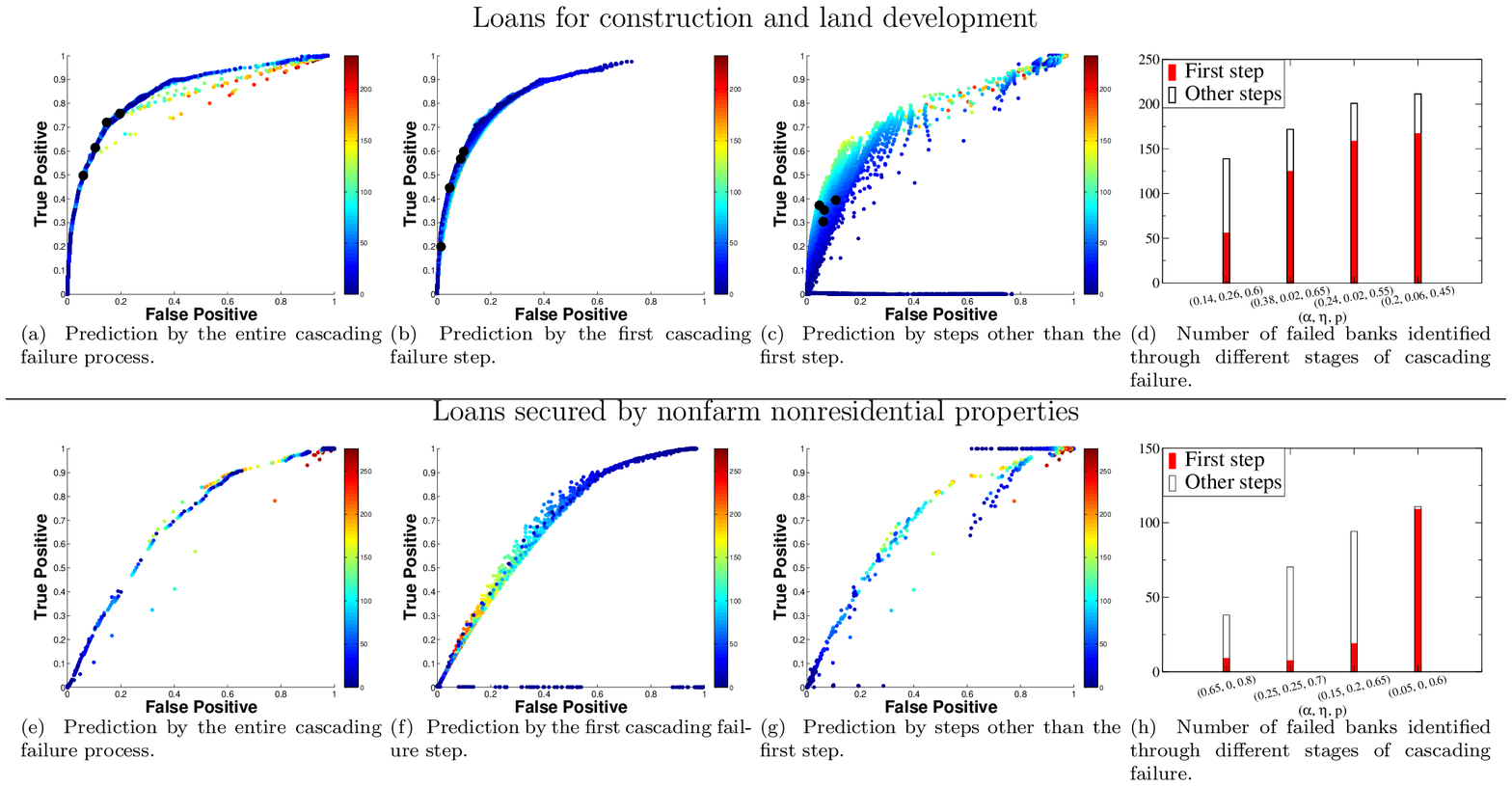}
  \caption{
  ROC curves of the prediction of failed banks by our cascading failure model when the loans for construction and land development are initially shocked (top figures) and when the loans secured by nonfarm nonresidential properties are initially shocked (bottom figures), based on 2007 data. Each point of the ROC curves corresponds to one combination of parameters $(\alpha, \eta, p)$. (a)(e) ROC curve of predictions made by the entire cascading failure process, (b)(f) of predictions made by the first cascading failure step and (c)(g) of predictions made by the other than the first cascading steps. The color of a dot represents the number of failed banks correctly identified by the model with the corresponding parameters combination. (d)(h) For fixed false positive rates of $5\%$, $10\%$, $15\%$, and $20\%$, we find parameter combinations with maximum true positive rates in fig.~(a), and show the number of failed banks identified by the first step (red) and the number of failed banks identified by the other steps in the cascading failure process (white). The black dots in (a)(b)(c) show the positions of four combinations respectively.
  }
  \label{roc_0}
\end{figure}

We firstly impose the initial shock to the construction and land
development loans and plot the ROC curves in the top row of fig.~\ref{roc_0}. 
As fig.~\ref{roc_0}(a) shows, when the false-positive rate is below
0.2 we have a relatively high true-positive to false-positive ratio.
For example, the four black dots in fig.~\ref{roc_0}(a) represent 
the false-positive rate and true positive rate pairs (0.06, 0.5), (0.1, 0.61), 
(0.15, 0.72) and (0.2, 0.78) respectively. The pair (0.06, 0.5) corresponds
to the parameter combination $(\alpha, \eta, p)=(0.14, 0.26, 0.6)$, which means
using this parameter combination, the model can identify $50\%$ of the actual 
failed banks that are on the FBL-FDIC with cost of $6\%$ false positive prediction. 
Overall, the ROC curve is
bended well above the diagonal curve, which means the model captures a
significant portion of the real-world behavior and has predictive power.

However, fig.~\ref{roc_0}(a) alone is not enough to justify
our complex networks model as necessary model to describe the systemic risk in
this banking system. 
If all of the actual failed banks owned a large amount of loans for construction 
and land development, then these banks will fail in the model in the first round
of failure after this type of asset is initially shocked. 
In that case, we only need to look at the weight of this asset in the banks' 
portfolio to identify the failed banks. 
However, we find that the failure of banks does not only occur because of the initial 
shock to specific assets, but also because of the amplified damage by positive 
feed back in the complex banking network. The interdependency between banks and the
complexity of network structure are crucial to this amplified damage in the system. 
To demonstrate our findings we conduct separately ROC curve analysis for the 
first-step prediction (bank failures caused directly by the initial shock on an asset) as well as for the 
consecutive-steps prediction (bank failures caused by a cascading failure process) 
as shown in figs.~\ref{roc_0}(b) and \ref{roc_0}(c).
We find that in addition to the first-step effective predictions, the 
consecutive-steps of the model further efficiently identify failed banks that can not be identified by
the first-step (ROC curve is above the diagonal line).  
Fig.~\ref{roc_0}(d) further shows the number of failed banks correctly
identified through the first and consecutive steps of the cascading failure
simulation for the four parameter combinations selected from fig.~\ref{roc_0}(a)
(black dots in the figures).
In all four cases, the number of failed banks predicted by the consecutive steps
represents a significant fraction of the total number of failed banks
identified. 
This result shows that some banks did fail only because of the the complex interconnections 
between banks in the system, which contributes to the risk contagion in the system. 
Thus, our model captures the complexity feature of the banking system and can offer
 prediction better than predictions made only based on balance sheet but without
 considering interactions between banks.

\begin{figure}[h!]
	\centering
	\includegraphics[width=\textwidth]{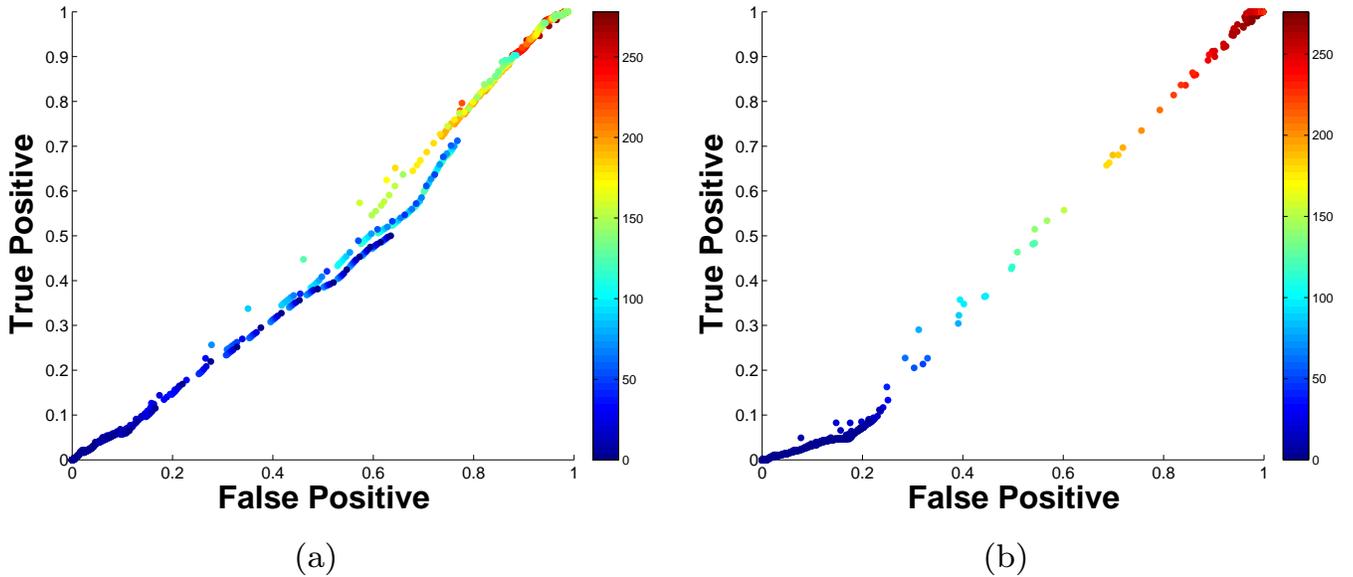}
	\caption{ROC curves of predictions of failed banks by our cascading failure model when (a) loans secured by 1-4 family resid. properties and (b) agricultural loans are initially shocked respectively. The straighter the ROC curve is, the closer it is to random case, meaning the less predictive power in regard to the failure of the commercial banks during the 2007-2011 financial crisis. 
}
\label{roc_other_assets}
\end{figure}

In addition to construction and land development loans, 
we also test our cascading failure model by simulating initial 
shock on other assets. The ROC curves in the bottom row of 
fig.~\ref{roc_0} show that the loans secured by nonfarm nonresidential 
properties, when initially shocked, have lower predictive power 
(smaller true-positive to false-positive ratio) compared to the case 
when initial shock is imposed on loans for construction and land development.
ROC curve tests for assets of loans secured by 1-4
family residential properties and agricultural loans, as shown in
figs.~\ref{roc_other_assets}(a) and \ref{roc_other_assets}(b), exhibit curves 
that are almost diagonal, indicating that initial shocks on these 
two assets have no predictive power on the failure of the banks in 
the 2007--2011 financial crisis. A truly random behavior would render 
points along the diagonal line (the so-called line of no discrimination) 
from the bottom left to the top right corners. 

The above ROC curve results suggest that the construction and
land development loans and the loans secured by nonfarm nonresidential
properties were the two asset types most relevant in the failure of
commercial banks during the 2007--2011 financial crisis.
It is largely believed that the past financial crisis is caused by 
residential real estate assets. However, we do not find evidence that loans 
secured by 1-4 family residential properties are responsible for 
commercial banks failures. 
This result is consistent with the conclusion of ref.~\cite{Cole2011} 
that the cause of the commercial banks failure between 2007-2011 were 
largely caused by commercial real estate-based loans rather than 
residential mortgages.

\begin{figure}[h!]
	\centering
    \includegraphics[width=0.7\textwidth ]{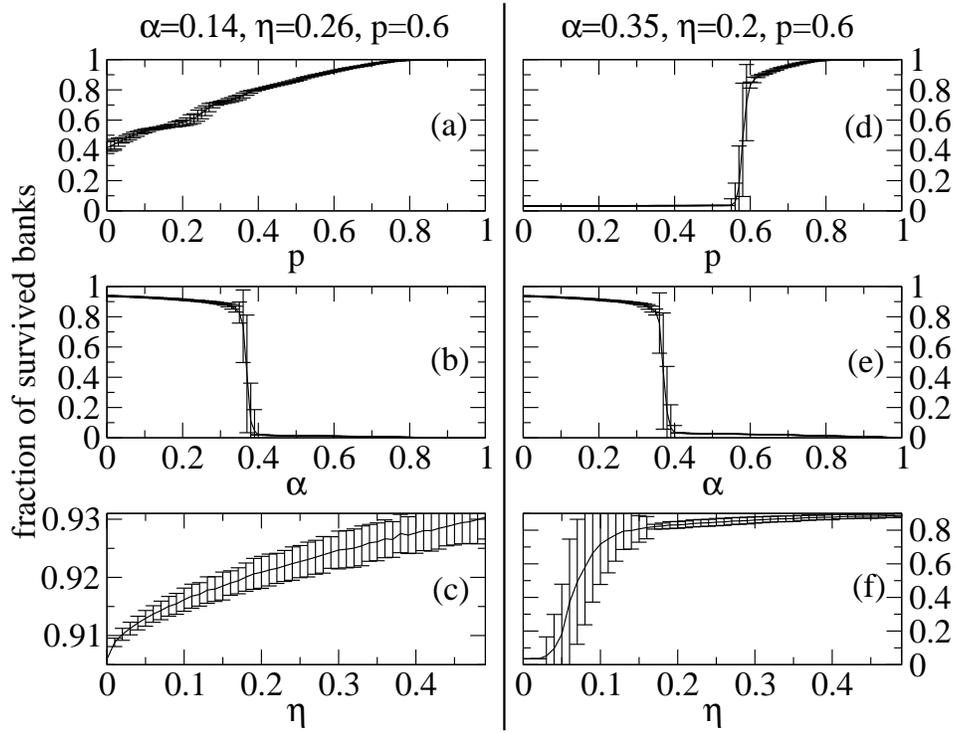}
	\caption{
	Survival rate of banks when asset 0 ( loans for construction and land development ) is initially shocked as function of one parameter with the other two parameters fixed. Average over 300 independent realizations with $95\%$ confidence interval. Left panel: parameter combination $\alpha=0.14$, $\eta=0.26$ and $p=0.6$; right panel: parameter combination $\alpha=0.35$, $\eta=0.2$ and $p=0.6$.
	}
	\label{parameter_impact}
\end{figure}

\begin{figure}[h!]
  \centering 	
  \includegraphics[width=0.8\textwidth ]{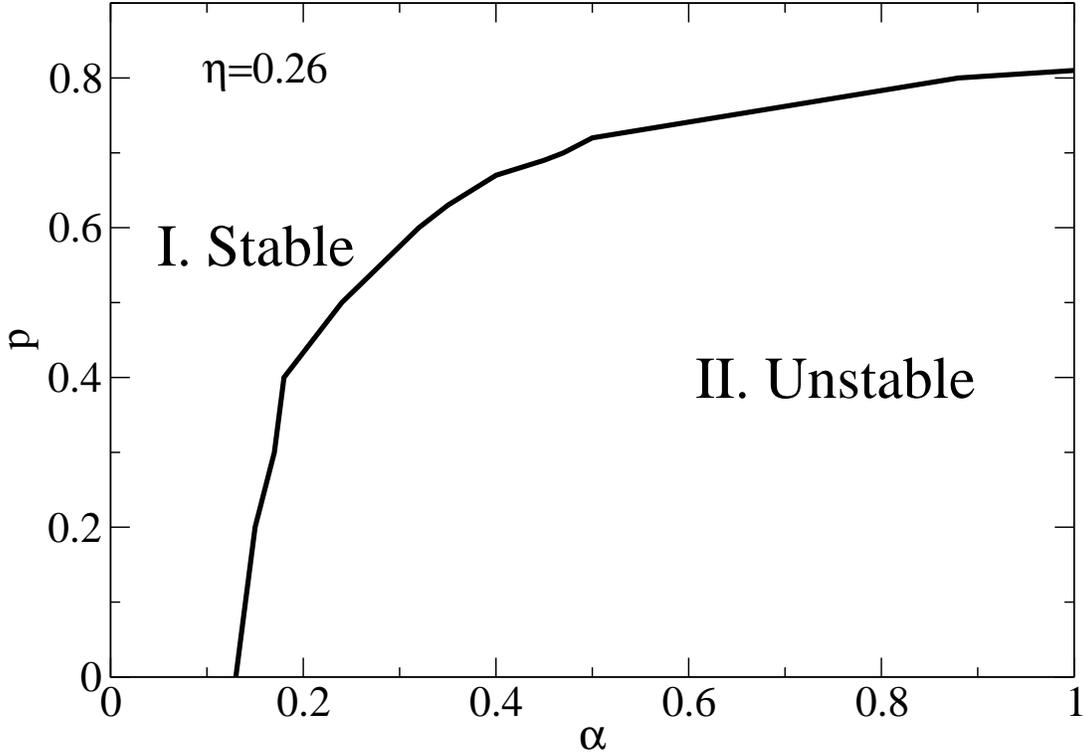}
  \caption{ Phase diagram for parameter alpha and p, when $\eta=0.26$. The network is stable in region I. Significant part of banks in the network would still survive after cascading failure. In region II, almost all the banks in the network fail after cascading failure.
  }
  \label{phase_diagram}
\end{figure}

Our final exploration is of the percolation-like property exhibited by
the bank-asset bipartite network. Complex networks usually exhibit
percolation phase transitions. As the dependent parameter changes, the
giant component of connected clusters in the network can drop to zero at
the critical point. In the bank-asset bipartite network model we go
beyond the giant component of connected clusters and study {\it all\/}
survived banks. Thus, percolation theory can not be applied. However, we
find that a percolation-like phenomenon also exists in this model. We
study the number of survived banks after the cascading failure process,
tuning one parameter and keeping the other two parameters fixed. We find
that the number of survived nodes in networks can change dramatically
with a small change of parameters. The parameter combination is chosen
as the first example in figure~\ref{roc_0}(d), $\alpha=0.14$, $\eta=0.26$, and
$p=0.6$. We show that the fraction of surviving banks changes smoothly
as parameters $p$ and $\eta$ change (see figs.~\ref{parameter_impact}(a) and
\ref{parameter_impact}(c)). But as $\alpha$ changes, the fraction of surviving
banks changes abruptly at a critical point and displays a
first-order-like abrupt phase transition (fig.~\ref{parameter_impact}(b)). We show that the first-order-like
phase transition also exists for $p$ and $\eta$ for a certain parameter
combination pool. As an example, we choose another parameter combination
($\alpha=0.35$, $\eta=0.2$, and $p=0.6$). We show in the right panel of
fig.~\ref{parameter_impact} that a first-order-like phase transition exists for all
three parameters, which means the system is at risk of abrupt
collapse. Figure~\ref{parameter_impact}(d) shows that, when the initial shock
parameter $p$ for an asset is below a certain threshold, even if the
other asset market values are undamaged, almost all banks default
because the cascading failure of this single asset (construction and land
development loans) significantly affects the overall financial
system. Figure~\ref{parameter_impact}(e) shows that when the effect of bank
failures on asset market values is sufficiently large, the whole banking
system is at risk of collapse. Figure~\ref{parameter_impact}(f) shows that when
$\eta$ is large, i.e., when the bank distress barrier of default is more
relaxed, the robustness of the system improves significantly. Thus, the
bank-asset bipartite network behaves differently for different parameter
combinations. Figure~\ref{phase_diagram} plots the phase diagram for
this bank network. Two different regions exist for parameters $p$ and
$\alpha$. In region I, the bank network system is in a stable state,
i.e., after cascading failure a significant number of banks will still
survive. In region II, the cascading failure process contributes to the
collapse of the entire bank network. Given that the bank network as a
complex system exhibits these two distinct states, it is extremely
important that policy makers institute rules that will keep the banking
system in the stable region.

\section*{Discussion}

In this paper, we develop a bipartite network model for systemic risk
propagation and specifically study the cascading failure process in the banking
system.  We first study the properties of the defaulting banks
during the 2007--2008 financial crisis, and find that they differ from
the properties of the survived banks. We then construct a bipartite 
banking network that is composed of (i) banks on one hand and (ii) bank assets on the other. 
We also propose a cascading failure model
to simulate the crisis spreading process in banking networks. We
introduce a shock into the banking system by reducing a specific asset
value and we monitor the cascading effect of this value reduction on
banks and on other asset values. We test our model using 2007 balance sheet data by
identifying the empirically failed banks between 2008 and 2011, and find 
through ROC curve analysis that our model simulates well the crisis spreading
process and identifies a significant portion of the actual failed banks
from the FDIC failed bank database.

Furthermore, studying the cascading failure of banks step by step shows that 
the complex structure of the bank network indeed contributes to the
spreading of financial crisis, which makes a complex network model necessary 
in describing and predicting the behavior of the banking system. Thus, we suggest
that our model could be useful to stress test systemic risk of the banking system.
For example, we can stress test the model to predict  which banks could be in danger 
and how many banks could fail if the agricultural assets drop $20\%$ in value.
We then offer policy suggestions such as requirement to reduce exposure to agricultural 
loans or closely monitor vulnerable banks.
Then the model also indicates possible ways to mitigate the propagation of
financial crisis. From the model we know that risk in the banking system propagates 
bidirectionally between assets and banks. Suppressing propagation either way could be very helpful to mitigating such catastrophes. The first way is to provide liquidity to the market, thus when distressed banks need to sell assets, the market will not overreact. The second way to curb systemic risk contagion is to ensure that banks are solvent and have healthy balance sheets, i.e. no excess leverage, higher capital requirements, appropriate levels of liquid assets, etc. in order to be able to absorb shocks to the asset value. Possible measures could be to pay a periodic fee to a supervising institution during non-crisis periods in exchange for obtaining emergency liquidity, as proposed by Perotti et al.~\cite{Perotti2009}.

Lastly, we show that as the parameters of the system change
the bank network can switch between two distinct regions, stable and unstable,
which are separated by a so-called phase transition boundary. 
We suggest that the bank network be understood in complex system terms and
that its closeness to the phase transition boundary be diligently monitored
in order to forestall system failure.

We suggest that our model for systemic risk propagation might be applicable to other
complex systems, e.g., to study the effect of sovereign debt value deterioration
on the global banking system or to analyze the impact of depreciation or
appreciation of certain currencies on the world economy. 

\section*{Methods}

\noindent{\bf Data Sets And Explanations.}  We use two data sets in this
paper. The first is the Commercial Banks - Balance Sheet Data (CBBSD)
from Wharton Research Data Services~\cite{WRDSweb} for the time period 1/1/1976 to
12/31/2008, which contains the amounts of 13 specific assets and the
total assets, total liabilities, and total equities for each bank. We enumerate the
assets from 0 to 12 to simplify the problem and categorize the assets
into real estate loans, other loans, and other assets. These assets are
listed in Table~\ref{asset_name}. 
We study the data for the year 2007, which contains 7,846 US banks. 
All banks have total assets data, but 21,171 data spots out
of the total $7,846 \times 13=101,998$ data spots for specific assets
are blank.
For banks with complete data, it is confirmed that the total
asset value equals the sum of individual asset. The absent data causes
the sum of the individual assets to be lower than the total assets.
Furthermore, in some cases, the sum of the individual assets can be smaller than the
bank's total liabilities, which leads the banks to fail before any shock 
is introduced in the model.
Thus we need to ensure that the sum of the individual asset values is equal to
the total assets value, by allocating the difference between the total asset
and available individual assets to the missing assets.
If a bank has more than one missing asset, the distribution of the difference 
to the assets is proportional to the average amount of these assets on
the balance sheets of other banks.

The step-by-step methodology is described as follows:

\begin{enumerate}

\item For each bank $i$, we calculate the weight $w_{i,m} =
  \frac{B_{i}{m}}{B_i}$ of asset $m$ in the bank's portfolio.

\item We then calculate the average weight of each asset $\langle
  w\rangle_{m} = \frac{\sum_{i}w_{i,m}}{N}$, where $N$ is the total
  number of banks.

\item From the total asset and known specific assets, we calculate the
  total amount for the unknown assets, which is $(B_i-\sum_{\mbox{known assets}} B_{i,m})$. 
  We then distribute this total amount to each unknown asset by 
  their average weight ($\langle w\rangle_{m}$)ratios. For
  example, if a bank $i$ lacks data on asset $x$ and asset $y$, the
  amount of asset $x$ is calculated as $B_{i,x}=
  (B_i-\sum_{m\neq\text{x, y}}B_{i,m})\frac{\langle w\rangle_{x}
  }{\sum_{m=x,y}\langle w\rangle_{m} }$.

\end{enumerate}

The second dataset that we use is the Failed Bank List from the Federal
Deposit Insurance Corporation (FBL-FDIC)~\cite{FDICweb}, which shows that 371 banks
failed during the 1/1/2008 -- 7/1/2011 period and that only 27 banks failed during
the 2000--2007 period. We use this representative dataset to empirically
test our model for the 2008 financial crisis. Of the 371 banks in the
FBL-FDIC dataset, 278 banks are included in the Commercial Banks - Balance
Sheet Data dataset in 2007.

\section*{Acknowledgments}
We  wish to  thank the European Commission FET Open Project "FOC" 255987 and "FOC-INCO" 297149, ONR, DTRA, the European EPIWORK, MULTIPLEX and LINC projects,   DFG  ,  the  Next  Generation  Infrastructures (Bsik) and the Israel Science Foundation for financial support.

\section*{Author Contributions}
All authors conceived and designed the research. X. H. carried out the numerical experiments. X. H., I. V. and S. H. analyzed the simulation results. All authors wrote the manuscript.

\section*{Additional Information}
The authors declare no competing financial interests.

\newpage
 \begin{table}[htbp]
\begin{tabular}{|l|l|p{8cm}|l|l|}
  \hline
  & index &{Balance Sheet Asset Variables} & {Rows} & {$\langle
  w\rangle_{m}$} \\ \hline\hline
  \multirow{5}{*}{Real Estate Loans} 
  & 0 & Loans for construction and land development & 6139 &0.082 \\
  & 1 & Loans secured by farmland & 5932 &0.038 \\
  & 2 & Loans secured by 1-4 family residential properties & 7553 &0.167 \\
  & 3 & Loans secured by multifamily (\textgreater 5) residential properties & 5381 &0.013 \\
  & 4 & Loans secured by nonfarm nonresidential properties. & 7495 &0.150 \\ \hline
  \multirow{5}{*}{Other Loans} 
  & 5 & Agricultural loans & 5167 &0.041 \\
  & 6 & Commercial and industrial loans & 3117 &0.031 \\
  & 7 & Loans to individuals & 7504 &0.097 \\
  & 8 & All other loans & 7049 & 0.171 \\
  & 9 & Obligations (other than securities and leases) of states and
  political subdivision in the U.S. & 7559 &0.046 \\\hline 
  \multirow{3}{*}{Other Assets} 
  & 10 & Held-to-maturity securities & 5924 &0.003 \\
  & 11 & Available-for-sale securities, total & 3445 &0.004 \\
  & 12 & Premises and fixed assets including capitalized lease & 7751 &0.020 \\
  \hline
\end{tabular}
\caption{ Description of Commercial Banks - Balance Sheet Data (CBBSD)
from Wharton Research Data Services. The third column represents the number of available rows of data of each asset for the year 2007 before completion. The total number of banks in 2007 in the CBBSD is $7846$. $\langle w \rangle _m$ is the average asset weight of banks. }
\label{asset_name}
\end{table}

\newpage

\end{document}